\documentclass[twocolumn]{aastex63}
\usepackage{graphicx}
\graphicspath{{figure/}}
\usepackage{subfigure}
\usepackage{color, hyperref, epsfig}
\usepackage{apjfonts, natbib}
\usepackage{appendix}
\usepackage{amsmath}
\usepackage{float}
\usepackage{bm}

\newcommand{\lum}{erg\,s\ensuremath{^{-1}}}

\newcommand{\mgii}{Mg\,{\footnotesize II}}

\newcommand{\feii}{Fe\,{\footnotesize II}}

\newcommand{\tdust}{\ensuremath{T\mathrm{_{dust}}}}
\newcommand{\ldust}{\ensuremath{L\mathrm{_{dust}}}}
\newcommand{\lbol}{\ensuremath{L\mathrm{_{bol}}}}
\newcommand{\lbolp}{\ensuremath{L\mathrm{_{bol,peak}}}}

\newcommand{\msun}{\ensuremath{M_{\odot}}}

\newcommand{\kms}{\ensuremath{\mathrm{km~s^{-1}}}}
\newcommand{\mbh}{\ensuremath{M_\mathrm{BH}}}

\shorttitle{PS16dtm dust model}
\shortauthors{Jiang et al.}

\begin{document}

\title{\Large The Extraordinary Long-lasting Infrared Echo of PS16dtm Reveals an Extremely Energetic Nuclear Outburst}

\author[0000-0002-7152-3621]{Ning Jiang}
\affiliation{CAS Key laboratory for Research in Galaxies and Cosmology, Department of Astronomy, University of Science and Technology of China, Hefei, 230026, China; jnac@ustc.edu.cn}
\affiliation{School of Astronomy and Space Sciences,
University of Science and Technology of China, Hefei, 230026, China}

\author[0009-0007-1153-8112]{Di Luo}
\affiliation{CAS Key laboratory for Research in Galaxies and Cosmology, Department of Astronomy, University of Science and Technology of China, Hefei, 230026, China; jnac@ustc.edu.cn}
\affiliation{School of Physical Sciences, University of Science and Technology of China, Hefei, 230026, China}

\author[0000-0003-3824-9496]{Jiazheng Zhu}
\affiliation{CAS Key laboratory for Research in Galaxies and Cosmology, Department of Astronomy, University of Science and Technology of China, Hefei, 230026, China; jnac@ustc.edu.cn}
\affiliation{School of Astronomy and Space Sciences,
University of Science and Technology of China, Hefei, 230026, China}

\author[0000-0002-0077-2305]{Roc M. Cutri}
\affiliation{IPAC/Caltech, 1200 E. California Boulevard, Pasadena, CA 91125, USA}

\begin{abstract}
PS16dtm is one of the earliest reported candidate tidal disruption events (TDEs) in active galactic nuclei (AGNs) and displays a remarkably bright and long-lived infrared (IR) echo revealed by multi-epoch photometry from the Wide-field Infrared Survey Explorer (WISE). After a rapid rise in the first year, the echo remains persistently at a high state from July 2017 to July 2024, the latest epoch, and keeps an almost constant color. We have fitted the extraordinary IR emission with a refined dust echo model by taking into account the dust sublimation process. The fitting suggests that an extremely giant dust structure with a new inner radius of $\sim1.6$~pc and an ultra-high peak bolometric luminosity, i.e., $\sim6\times10^{46}$~\lum\ for typical 0.1~$\mu$m-sized silicate grain, is required to account for the IR echo. 
This work highlights the distinctive value of IR echoes in measuring the accurate intrinsic \lbol, and thus the total radiated energy of TDEs, which could be severely underestimated by traditional methods, i.e. probably by more than an order of magnitude in PS16dtm. 
Such large energetic output compared to normal TDEs could be boosted by the pre-existing accretion disk and gas clouds around the black hole. Our model can be validated in the near future by IR time-domain surveys such as Near-Earth Object (NEO) Surveyor, given the recent retirement of WISE. In addition, the potential for spatially resolving a receding dusty torus after a TDE could also be an exciting subject in the era of advanced IR interferometry.
\end{abstract}


\section{Introduction}

The discovery and study of stellar tidal disruption events (TDEs) by supermassive black holes (SMBHs) have made great progress thanks to wide-field time-domain surveys over the last decade~\citep{Gezari2021}. In particular, the Zwicky Transient Survey (ZTF) has boosted the discovery rate of optical TDEs from $\lesssim$2/yr to $>$10/yr~\citep{vV2021,Hammerstein2023} and the eROSITA has provided the first X-ray TDE sample from a single uniform X-ray survey~\citep{Sazonov2021}. These surveys mark a new era in TDE sample statistics, allowing an unprecedented demography of dormant SMBHs~\citep{Yao2023}. Traditionally, TDE selection has excluded active galactic nuclei (AGNs), focusing solely on inactive galaxies (e.g., \citealt{vV2021}). However, in theory, TDE rates in AGNs should be significantly higher compared to normal galaxies (e.g., \citealt{Karas2007,Kennedy2016,Kaur2024,Wang2024}). Observations have indeed identified a growing number of TDE candidates in AGNs (e.g., \citealt{Blanchard2017,Kankare2017,Trakhtenbrot2019,Liu2020,Zhang2022}), although their TDE nature has not been definitively distinguished from other imposters~\citep{Zabludoff2021}.

A notable characteristic of TDEs in AGNs is the prominent infrared (IR) echoes they produce, namely the reprocessed emission of dust in the vicinity of SMBHs~\citep{Lu2016,Jiang2016,vV2016} due to the presence of a dusty torus in AGNs.  Their peak IR luminosity and inferred dust covering factor on subparsec scale, defined as the ratio between the IR luminosity and optical bolometric luminosity, are at least one order of magnitude higher than normal optical TDEs~\citep{Jiang2021b}. It is notable that a newly discovered population of IR-selected TDEs in inactive galaxies shows intense IR echoes with a luminosity comparable with that of TDEs in AGNs~\citep{Jiang2021a,Wang2022,Panagiotou2023,Masterson2024}. Therefore, the IR echoes of TDEs and other ambiguous nuclear transients provide us a new opportunity to probe the torus structure (e.g., \citealt{Dou2017,Jiang2019}) or, more generally, sub-parsec dust environments of SMBHs (e.g., \citealt{Jiang2021b,Hinkle2024,Newsome2024}). Furthermore, they can also serve as bolometers to measure the intrinsic luminosity and released energy of TDEs by observing the reprocessed emission of high-energy photons~\citep{vV2016}, including the unobserved extreme ultraviolet (EUV) photons~\citep{Dai2018}. 

\begin{figure*}
\centering
\begin{minipage}{0.9\textwidth}
\centering{\includegraphics[width=1.0\textwidth]{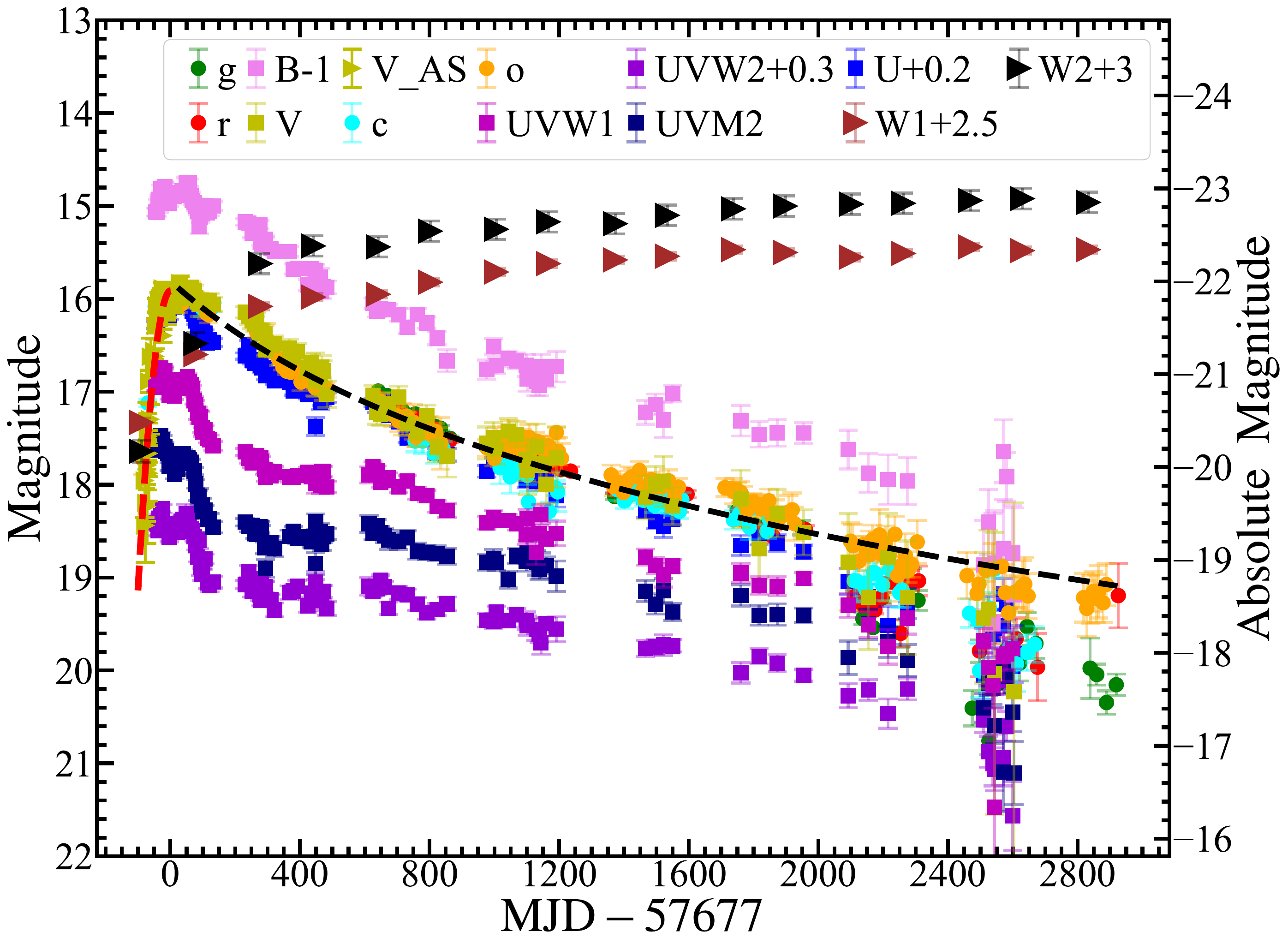}}
\end{minipage}
\caption{The multiwavelength light curves for PS16dtm. The Swift/UVOT photometry is denoted by square in different colors, the ZTF and ATLAS photometry by dots and the WISE photometry by triangles. The red dashed line shows the estimated Gaussian-fitted rise of PS16dtm using both ASASSN-V and Swift-V ($t_{\rm peak}=57677\pm5, \sigma_{\rm rise}=58\pm2$). The black dashed line shows the estimated power-law decay fitted decline of PS16dtm using ATLAS-o band with an index of $-1.66\pm0.06$. All magnitudes are in the AB magnitude system, except for WISE, which uses the Vega magnitude system.} 
\label{olc}
\end{figure*}

In this letter, we present a comprehensive analysis of the extraordinary IR echo in PS16dtm, one of the earliest proposed TDEs in AGNs, and explore its value in revealing the structure of the torus and the intrinsic bolometric luminosity of the TDE. PS16dtm was first reported by \citet{Blanchard2017} as a TDE in the nucleus of a known narrow-line Seyfert~1 (NLSy1) galaxy, a class of AGNs that preferentially exhibit rapid enhanced flaring activity~\citep{Frederick2021}. Soon after, \citet{Jiang2017} reported the discovery of its early IR echo, with the first IR detection 11 days earlier than the first optical detection. In the first epoch, the inferred blackbody temperature was even higher than the dust sublimation temperature of common grains, indicating an ongoing sublimation process. The evaporated dust can naturally explain the rapid emergence of prominent \feii\ complexes as the metals initially locked in the dust are released ~\citep{Jiang2017,He2021}. Extensive multi-wavelength follow-up of PS16dtm, including photometry and spectroscopy in the UV/optical, has been carried out that spans ~2000 days~\citep{Petrushevska2023}. These observations show that the properties of PS16dtm are difficult to reconcile with normal AGN variability due to its much larger amplitude and shorter rise timescale ($<$100 days). Similarly, its stable color and distinctive spectroscopic evolution disfavor a supernova origin. In particular, \citet{Blanchard2017} argued that neither the supernova scenario nor AGN variability can account for the strange X-ray dimming. Instead, they suggested that this phenomenon could be explained by stellar debris that obscures the X-ray-emitting region of the AGN accretion disk. Therefore, the TDE scenario remains the most plausible explanation. PS16dtm was also selected in the sample of mid-infrared outbursts in nearby galaxies (MIRONG, \citealt{Jiang2021a}), which found that its MIR light curves continue to rise until 2019. \citet{Petrushevska2023} also noticed its long-lived IR echo with a rising trend that persists until July 2021, but without any explanatory model.

We assume a cosmology with $H_{0} =70$ km~s$^{-1}$~Mpc$^{-1}$, $\Omega_{m} = 0.3$, and $\Omega_{\Lambda} = 0.7$.

\section{Data}
\subsection{Optical and UV light curves}
We have collected all archival data from the Ultraviolet/Optical Telescope (UVOT; \citealt{Roming2005})
on board the Neil Gehrels Swift Observatory (hereafter Swift). The data were reduced following the standard pipeline of \texttt{Heasoft}. We then ran \texttt{uvotimsum} to sum the images and then generated the light curves using the \texttt{uvotsource} task, with the source and background regions defined by circles with radii of $5^{\prime\prime}$ and $30^{\prime\prime}$ respectively. We also collected the forced photometric light curves of PS16dtm from the Asteroid Terrestrial Impact Last Alert System (ATLAS; \citealt{Tonry2018,Smith2020}), the Zwicky Transient Facility (ZTF; \citealt{Masci2019}) and the All Sky Automated Survey for SuperNovae (ASAS-SN, \citealt{Shappee2014,Kochanek2017}). The forced measurements of the ATLAS,ASAS-SN and ZTF surveys have already been performed on the difference images. Since ATLAS and ASAS-SN began their observations prior to the occurrence of PS16dtm, the flux contribution from the host galaxy could be directly subtracted from the reference images. However, for the ZTF data, the flux values are negative because the observations only cover the declining phase, with the reference images sampling the period of a higher phase, and the light curve has not yet decayed to a quiescent phase. So we performed PSF photometry on the reference images and added this constant flux to the ZTF light curves first. Then for the ZTF and Swift/UVOT data, we subtracted the PS16dtm host galaxy magnitudes from \citet{Hinkle2021}, which have performed a host galaxy SED fit for PS16dtm. All of these photometric measurements have been converted to the AB magnitude system and are presented in Figure~\ref{olc} after correction for Galactic extinction.

To investigate the physical parameters of the optical/UV (OUV) emission of PS16dtm, we performed a blackbody fit to the SEDs at different epochs. Our results are consistent with those of \citet{Blanchard2017} and \citet{Petrushevska2023}, indicating that the SEDs of PS16dtm cannot be adequately described by a single blackbody due to notable suppression in the UV bands.
Consequently, we also applied blackbody fitting to our photometric data, excluding the NUV bands, to estimate the bolometric light curves, requiring at least four photometric measurements at each epoch.
We set the rest-frame reference date as the moment of maximum blackbody luminosity, corresponding to $\rm MJD\sim57653$ in the following analysis.

\subsection{Mid-IR light curve}

\begin{figure}
\centering
\begin{minipage}{0.5\textwidth}
\centering{\includegraphics[width=0.9\textwidth]{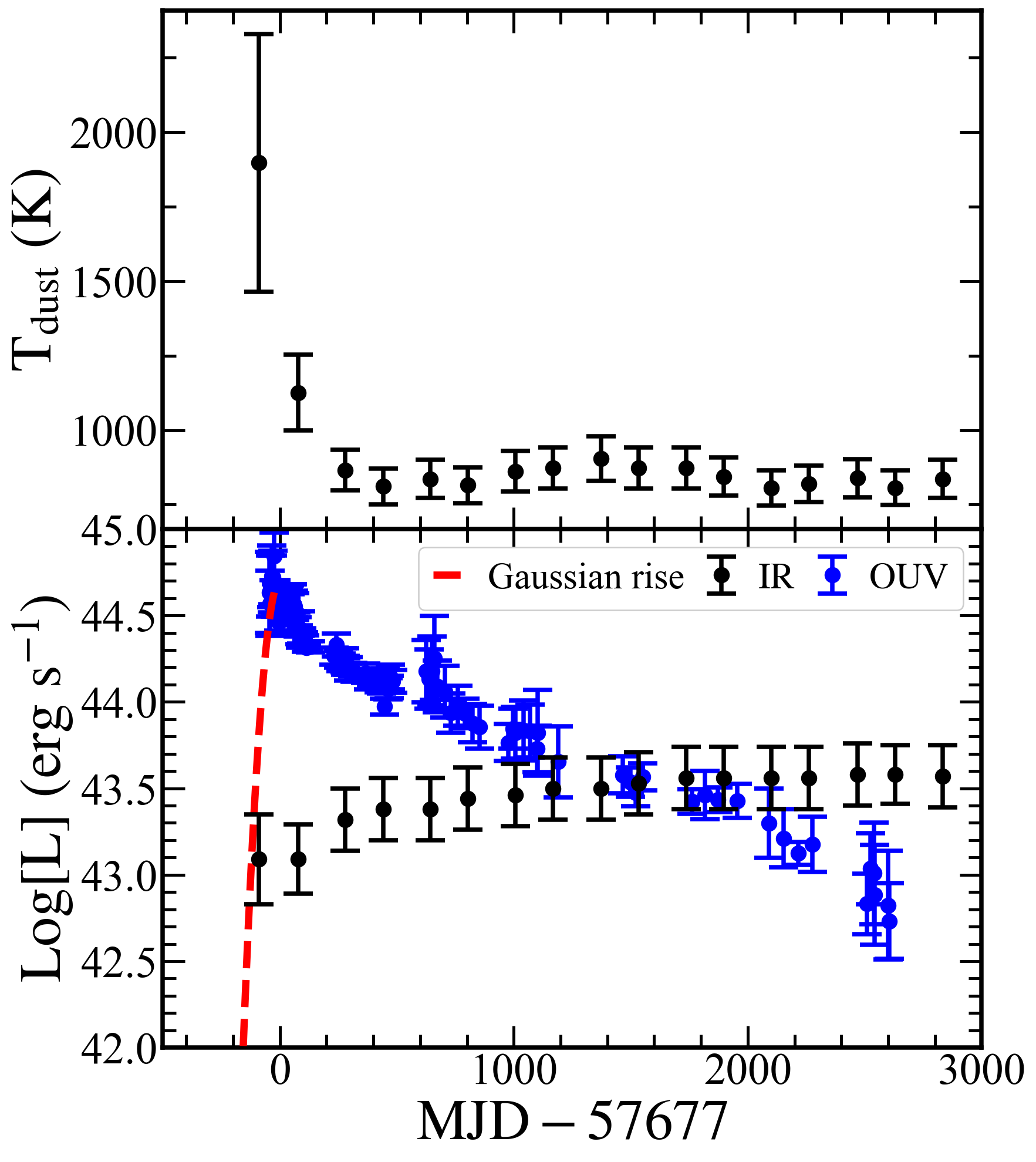}}
\end{minipage}
\caption{The dust temperature (top panel) and luminosity (bottom panel) evolution of PS16dtm. In the bottom panel, the black and blue dots denote the dust IR luminosity and OUV luminosity, respectively. The red dashed line corresponds to the rising phase of the OUV light curve, which is fitted with a Gaussian function. This phase is poorly sampled by observations but is essential for our model.}
\label{dust}
\end{figure}

Following our previous works~\citep{Jiang2017,Jiang2021a}, we retrieved and binned the multi-epoch mid-IR (MIR) photometry at the position of PS16dtm retrieved from Single-Exposure photometry databases of the Wide-field Infrared Survey Explorer (WISE, \citealt{Wright2010}) and its follow-on mission, the Near-Earth Object WISE (NEOWISE, \citealt{Mainzer2014}). They provide us with a long-term light curve that spanned the period from January 14, 2010, to July 19, 2024, with a half-year cadence. This is the final available WISE light curve because the mission ended shortly after the last observation epoch. Following the procedure described in~\citet{Jiang2021a}, we first filtered the data using the quality flags marked in the catalogs. The bad data points with poor-quality frame (\texttt{qi\_fact}<1),
charged particle hits (\texttt{saa\_sep}<5), scattered moonlight (\texttt{moon\_masked}=1)
and artifacts (\texttt{cc\_flags}$\neq$0). The remaining data are then binned every half year to increase the signal-to-noise ratio of the photometry.
We established the baseline quiescent level by averaging the data before the outburst and then calculated the magnitudes associated with PS16dtm with this baseline subtracted. A total of 17 epochs of echoes have been detected up to the end of the observations in July, with no obvious decreasing trend (see Figure~\ref{olc}).

We computed the dust properties, specifically its temperature (\tdust) and luminosity (\ldust), as in \citet{Jiang2021a}, which assumes a silicate dust grain composition and takes into account the dust absorption coefficient. The evolution of \tdust\ and \ldust\ is shown in Figure~\ref{dust}. Note that although \tdust\ is systematically lower than that given in \citet{Jiang2017} after considering the absorption coefficient, it is still close to the silicate sublimation temperature in the first epoch ($\tdust=1897\pm432$~K).  \tdust\ remains relatively stable from the third epoch onward, with an average value of 844~K. It becomes slightly higher (about 980 K) when fitted with a blackbody model, while \ldust\ shows very negligible differences.

\section{Analysis}
\subsection{Dust echo model}
\label{model}

\begin{figure*}
\centering
\begin{minipage}{1.0\textwidth}
\centering{\includegraphics[width=1.0\textwidth]{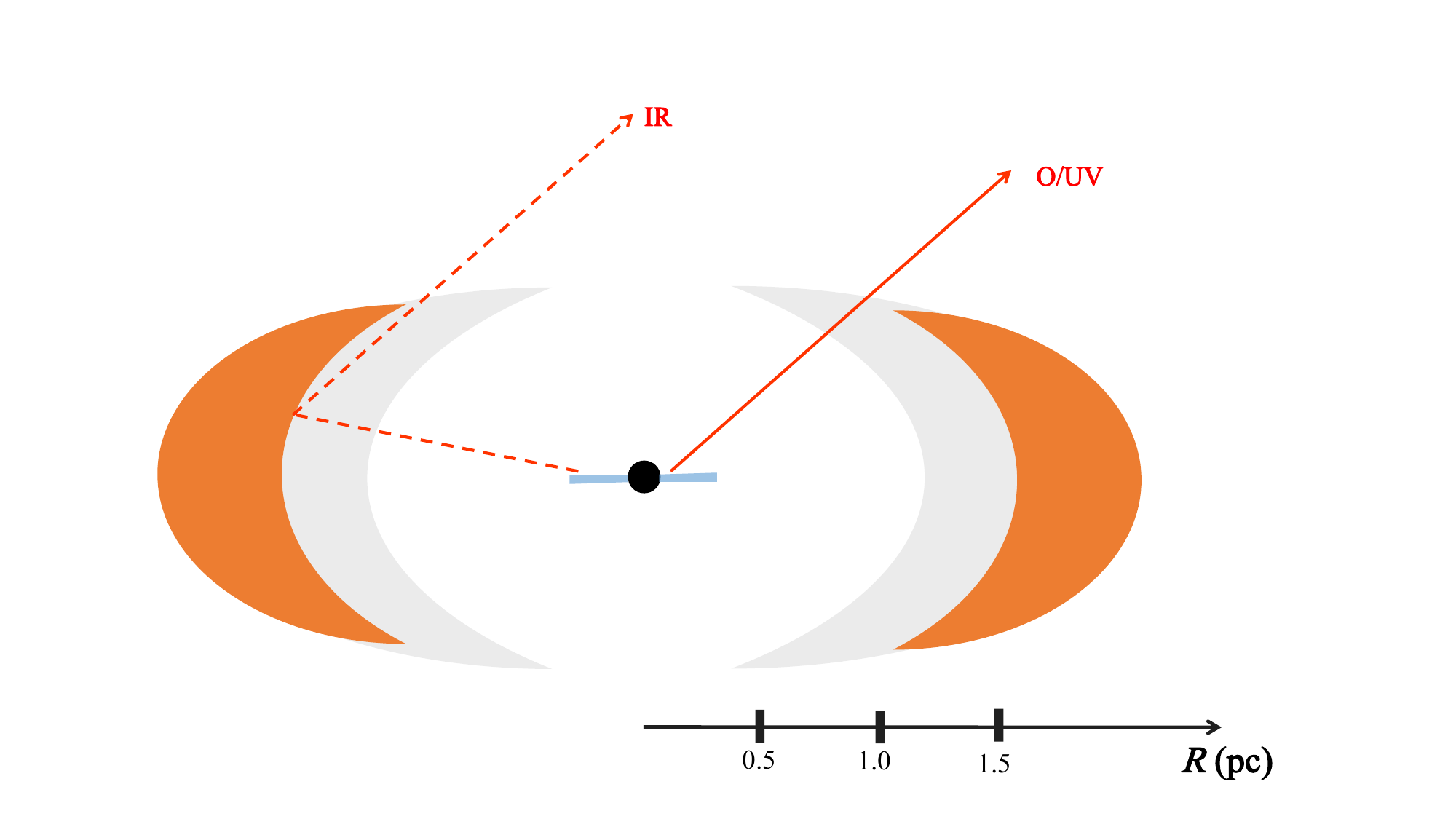}}
\end{minipage}
\caption{Schematic picture of the dust echo model. The central SMBH, accretion disk and dusty torus are shown in black, blue and orange, respectively. The grey region represents the evaporated part after the outburst of PS16dtm. We have also marked the physical scale at the bottom to show the evaporation area of the torus. Note that the scale does not apply to the BH and accretion disk, which are otherwise too small to visualize. }
\label{pic}
\end{figure*}

The dust echo model is commonly used to describe how the IR emission from the torus responds to the OUV variability of an AGN (see Figure~\ref{pic}). 
Observations show that the quantity $2\Delta T_{\text{IR}}$ is closely related to the IR time delay, representing the interval between the peaks of IR and OUV emission in the SMBH frame. This parameter helps to derive the approximate radius of the inner edge of the dust torus, which can be estimated as $R_{\text{dust}}\approx c\Delta T_{\text{IR}}$. The dust echo model provides a predictive framework for the IR luminosity light curve given by:

\begin{equation}
L_{\text{IR}}(t) =\epsilon_\Omega \epsilon_{\text{dust}} \int_{-\infty}^{+\infty} L_{\text{OUV}}(t') f(t-t') \mathrm{d}t'
\label{lir}
\end{equation}

Here, $L_{\text{OUV}}$ is the OUV luminosity, $\epsilon_\Omega$ is the solid angle covering factor of the dust distribution, $\epsilon_{\text{dust}}<1$ is the fraction of incident radiation that is reprocessed into IR radiation by the illuminated dust, and $f(t-t')$ is the time spreading function. Previous studies (e.g.,~\citealt{vV2016,Yuan2024}) usually employ a box function for $f(t-t')$, assuming a static spherical dust shell. However, the high temperature inferred from the IR emission immediately after the outburst suggests a dust sublimation process at the inner edge of the dust shell. This requires a refined definition of $f(t-t')$ to include the effect of the dynamical evolution of the dust radius, as follows:

\begin{equation}
f(t-t') =
\begin{cases}
    \frac{c}{2R_{\text{dust,i}}} &t'\le t_{\text{ini}}\ \&\ 0<t-t'<\frac{2R_{\text{dust,i}}}{c} \\
    \frac{c}{2R_{\text{dust,sub}}} &t_{\text{ini}}<t'<t_{\text{peak}}\ \&\ 0<t-t'<\frac{2R_{\text{dust,sub}}}{c} \\
    \frac{c}{2R_{\text{dust,f}}} &t'\ge t_{\text{peak}}\ \&\ 0<t-t'<\frac{2R_{\text{dust,f}}}{c} \\
    0 &\text{otherwise}
\end{cases}
\label{timespread}
\end{equation}
where $R_{\text{dust,i}}$ is the initial radius, $t_{\text{ini}}$ is the rest-frame time when the dust started to sublimate, $R_{\text{dust,f}}$ is the final radius, $t_{\text{peak}}$ is the rest-frame time when the light from the peak of the transient flare reached the inner edge of the dust shell, and
\begin{equation}
R_{\text{dust,sub}} =R_{\text{dust,f}} \left(\frac{L_{\rm bol}}{\lbolp}\right)^{1/2}
\label{rsub}
\end{equation}
is the radius during the sublimation. Here, \lbolp\ is the peak bolometric luminosity. This relationship is based on the fact that the inner radius is directly proportional to the square root of the bolometric luminosity when the dust evaporation time scale is negligible.

We have ignored the dust scattering in our model, which may contribute a non-negligible fraction to the tail of the optical light curve, for two reasons. 
First, the tail has little impact on our fitting results, since the IR light curve remains in the high phase, and thus the dust emission is predominantly reprocessed by the peak OUV emission. Second, the dust covering factor inferred from our previous work is $\sim0.1$~\citep{Jiang2021b} and the real value should be even smaller due to an underestimation of the OUV luminosity and supported by the low $\epsilon_\Omega \epsilon_{\text{dust}}$ value in our fit (see \S\ref{modelfit}). 

\subsection{Model fitting}
\label{modelfit}

We assume that the observed luminosity ($L_{\text{bol,obs}}$) of PS16dtm derived from traditional blackbody fitting is a constant fraction of the real intrinsic \lbol, that is
\begin{equation}
L_{\text{bol}} \simeq L_{\text{OUV}} =\lambda L_{\text{bol,obs}}
\label{lam}
\end{equation}
where $\frac{1}{\lambda}<1$ is the fraction.

To perform a fit to the optical and IR light curves using the dust echo model described in \S\ref{model}, we employ a likelihood function as:
\begin{equation}
\mathcal{L} \varpropto \prod_{i=1}^N \left[\frac{1}{\sqrt{2\pi}\sigma_{\text{peak}}}e^{-\frac{(L_{\text{peak,1}}-L_{\text{peak,2}})^2}{2\sigma_{\text{peak}}^2}}
\frac{1}{\sqrt{2\pi}\sigma_{\text{IR}}}e^{-\frac{(L_{\text{IR,obs}}-L_{\text{IR}})^2}{2\sigma_{\text{IR}}^2}}\right]
\label{lf}
\end{equation}
where $L_{\text{IR,obs}}$ is the observed IR luminosity and $\sigma_{\text{IR}}$ is the associated error, $L_{\text{IR}}$ is the luminosity at the corresponding time computed by the dust echo model (See Eq.~\ref{lir}), $L_{\text{peak,1}}$ is the peak luminosity derived from Eq.~\ref{lam}, and $L_{\text{peak,2}}$ is derived from the relation below~\citep{Namekata2016,Jiang2017}:
\begin{equation}\label{lpeak}
    \frac{L_{\rm bol,peak}}{10^{45}~\rm erg~s^{-1}}=\left( \frac{R_{\rm sub}}{\rm 0.121~pc} \right)^2 \left(\frac{T_{\rm sub}}{\rm 1800~K} \right)^{5.6} \left(\frac{a}{0.1\mu\rm m} \right)
\end{equation}
where $R_{\rm sub}$ is the sublimation radius, $a$ is the grain size, and $T_{\rm sub}$ is the dust temperature.

\begin{figure*}
\centering
\begin{minipage}{1.0\textwidth}
\centering{\includegraphics[width=1.0\textwidth]{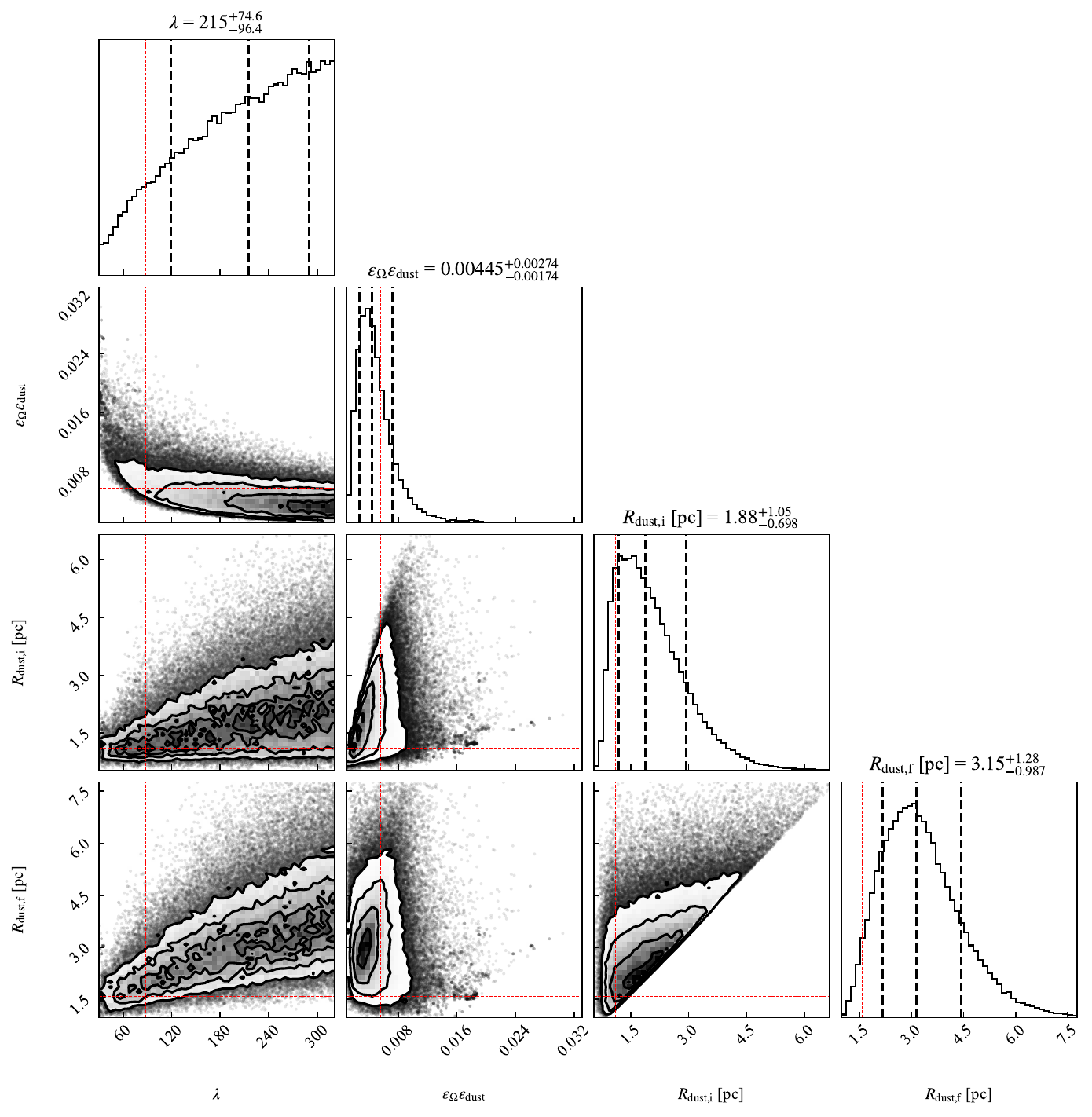}}
\end{minipage}
\caption{The distribution of dust echo model parameters for PS16dtm. The unit of $R_\text{dust,i}$ and $R_\text{dust,f}$ is light year. The red dashed lines denote the location of maximum likelihood, corresponding to the parameter values of $\lambda=87.7$, $\epsilon_\Omega \epsilon_{\text{dust}}=0.0056$, $R_{\text{dust,i}}=1.10$~pc, and\ $R_{\text{dust,f}}=1.59$~pc. The values and errors above the plots correspond to the $15.87\%$, $50\%$ and $84.13\%$ quantile values of the posterior samples of the parameters, and the black dashed lines show the locations of these values in the plots. The contours in the 2D contour plots reflect the relative numerical density, where darker color means a larger numerical density.}
\label{para}
\end{figure*}

For subsequent simulations, we set $T_{\rm sub}=1500$~K and $\sigma_{\text{peak}}=\sqrt{2}$ dex. This gives parameter values of $\lambda=87.7$, $\epsilon_\Omega \epsilon_{\text{dust}}=0.0056$, $R_{\text{dust,i}}=1.10$ pc, and\ $R_{\text{dust,f}}=1.59$ pc.
The uncertainties from the choice of $T_{\rm sub}$ and $\sigma_{\text{peak}}$ have some impact on the simulation results, but not substantially. For example, if we set $T_{\rm sub}=1200$~K and $\sigma_{\text{peak}}=\sqrt{2}$ dex, we would get $\lambda=44.2$, $\epsilon_\Omega \epsilon_{\text{dust}}=0.0111$, $R_{\text{dust,i}}=1.10$ pc, and\ $R_{\text{dust,f}}=2.10$ pc; if we set $T_{\rm sub}=1500$~K and $\sigma_{\text{peak}}=1$ dex, we would get parameter values of $\lambda=109.6$, $\epsilon_\Omega \epsilon_{\text{dust}}=0.0044$, $R_{\text{dust,i}}=1.09$ pc, and\ $R_{\text{dust,f}}=1.78$ pc. All these values fall in the same parameter space depicted in Figure~\ref{para}, i.e., an accurate model setup would not be significantly helpful in constraining the posterior parameter samples.

To establish a continuous $L_{\text{OUV}}$ for calculating simulated IR luminosities, it is assumed that the $L_{\text{OUV}}$ between two data points follows a linear trend. We use the estimated \lbol\ of AGN from~\citet{Jiang2017}, i.e., $10^{42.73}$~\lum, to represent $L_{\text{OUV}}$ before the flare. Furthermore, we assume that $L_{\text{OUV}}$ of PS16dtm will decay exponentially in the near future. The IR emission from dust sublimation should be described by a more sophisticated physical model than the simple dust echo model presented here. In this work, we focus mainly on the inner radius of the dust shell, which could be roughly described by our simple model. Therefore, our simulation ignores the first and second IR data points to avoid getting stuck in the dust sublimation period.

\begin{figure}
\centering
\begin{minipage}{0.5\textwidth}
\centering{\includegraphics[width=0.9\textwidth]{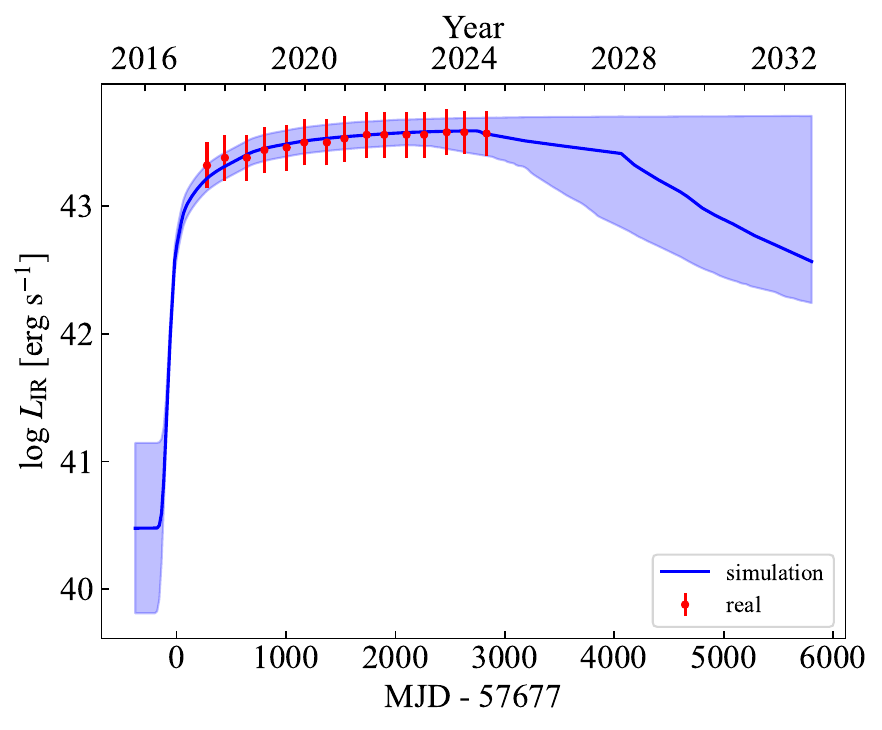}}
\end{minipage}
\caption{The best-fitting IR luminosity light curve. The red errorbars denote the real IR luminosity estimated from black body fit of WISE data, while the blue curve denote the simulated IR light curve. The shaded blue area show the 3$\sigma$ uncertainty of the simulated light curve.}
\label{bestir}
\end{figure}

We set a uniform prior for the parameters $\lambda$, $\epsilon_\Omega \epsilon_{\text{dust}}$, $R_{\text{dust,i}}$ and $R_{\text{dust,f}}$. We apply the Markov Chain Monte Carlo (MCMC) sampler \texttt{emcee}~\footnote{https://emcee.readthedocs.io/en/stable/}~\citep{Foreman-Mackey2013} to construct the posterior samples of the parameters. The distributions of the posterior samples are shown in Figure~\ref{para}.
We also show the best-fitted IR luminosity light curve in Figure~\ref{bestir}. To assess the uncertainties, we randomly selected 10,000 parameters from the posterior samples and performed a statistical analysis on their corresponding simulated light curves. The model predicts that MIR emission will begin to decline rapidly after 2027. However, the uncertainty of the turning point time could be as large as several years.

By substituting the measured $R_{\rm sub}$ of 1.59~pc and a $T_{\rm sub}$ of 1500~K for silicate grains in Eq.~\ref{lpeak}, we obtain $L_{\rm bol}=6.2\times10^{46} a_{\rm 0.1\mu m}~{\rm erg~s^{-1}}$. The \lbolp\ depends significantly on the grain size, which can vary from $6.2\times10^{45-47}$~\lum\ for the size $0.01-1\mu$m. Assuming an average grain size of 0.1~$\mu$m, the corresponding \lbolp\ is $6.2\times10^{46}$~\lum. Given a black hole mass (\mbh) of $10^6$~\msun~\citep{Xiao2011,Blanchard2017}, we obtain an Eddington ratio of about $4.8\times10^2$, suggesting a highly super-Eddington accretion system. Our calculated \lbolp\ is $\sim$ 90 times higher than the blackbody luminosity fitted from the optical SED, and we will discuss its implications in \S\ref{tde-sed}.

\section{Discussion}
                                                        
\subsection{Implication to the intrinsic SED of TDEs and the missing energy puzzle}
\label{tde-sed}

Our dust echo strongly suggests a highly super-Eddington accreting system for PS16dtm, although the specific Eddington ratio depends on the dust grain properties. In fact, a super-Eddington accretion phase is well predicted for a solar-like star disrupted by a $10^{6}$~\msun\ BH from early theoretical calculations~\citep{Rees1988,Evans1989} and recent modern simulations~\citep{Bandopadhyay2024}.
Our result indicates that the blackbody assumption may have vastly underestimated the real \lbol. In fact, the intrinsic SED of a TDE cannot be directly probed because extreme UV photons, i.e., from $\sim40$~\AA\ to $\sim1000$~\AA, are unobservable due to attenuation along the line of sight by neutral hydrogen and dust. The underestimation of \lbol\ in PS16dtm may be even more complicated, as NUV photometry is significantly lower than expected from the blackbody fit, indicating likely a strong additional dust extinction effect, i.e., by circumnuclear dust near the SMBH (see Figure~5 in \citealt{Blanchard2017}). \citet{Petrushevska2023} also noticed this problem and tried to fit a single blackbody to the NUV/optical data, assuming an extinction similar to that measured in the Small Magellanic Cloud (SMC). They found that the fitted blackbody has an unreasonably high temperature of $\sim (2 - 4)\times10^5$ K and a \lbol\ of $10^{48} - 10^{49}$~\lum, concluding that a single blackbody cannot provide a satisfactory fit to the SED, even when extinction is included. 

\begin{figure*}
\centering
\begin{minipage}{1\textwidth}
\centering{\includegraphics[width=0.9\textwidth]{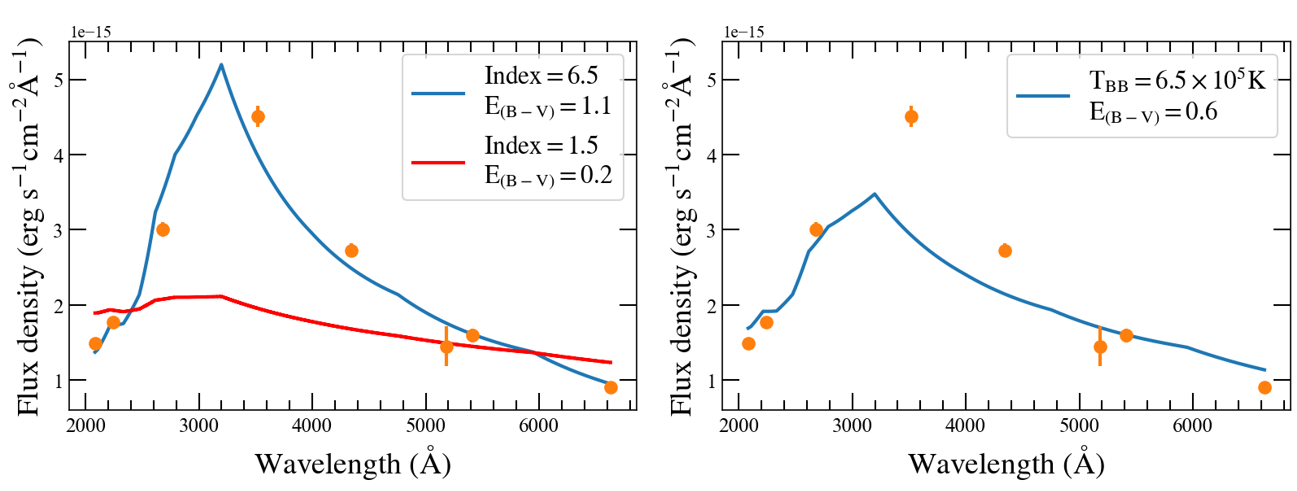}}
\end{minipage}
\caption{The single power-law (left panel; $f=A\lambda ^{-n}$) and blackbody (right panel) fits to the SED of PS16dtm around its optical peak ($\rm MJD\sim57730.5$, +72\,d). The fitting incorporates a free parameter for dust extinction, modeled similarly to that observed in the SMC. In each panel, the best-fit SED is represented by a blue line, while the orange dots denote the observed photometry at the peak epoch. The power-law fit yields a very steep slope (index $n=6.5$) and a $\chi^2$ value of 10.9, compared with an AGN-like SED (index fixed at $n=1.5$) fitting result with $\chi^2$ of 141.2,  and the blackbody fit gives a very high temperature ($6.5\times10^5$~K) yet with a much worse $\chi^2$ value of 47.7. It is worth noting that the strong extinction from the SMC introduces some of the bumps and wiggles observed in the otherwise smooth model.}
\label{SED_fit}
\end{figure*}

We have also checked the peak SED of PS16dtm and found that neither a dust-extincted blackbody nor an AGN-like power-law SED  ($f=A\lambda ^{-n}$ with $n$ fixed to a typical value of 1.5) provides a satisfactory fit. In contrast, an extremely steep ($n=6.5$) power-law function with a high dust extinction ($E_{\rm (B-V),\ host}=1.1$) seems to match observations much better (see Figure~\ref{SED_fit}). We emphasize that the dust extinction effect has already been accounted for in our dust echo modelling by the parameter $\lambda$ (\S\ref{modelfit}), which is a total underestimation factor due to both dust extinction and an incorrect SED assumption. Such a steep power-law OUV SED has actually been predicted by a unified reprocessing model, in which the optically thick outflow launched from the radiation-pressure-dominated thick accretion disk during the super-Eddington phase naturally serves as the role of the reprocessing layer in the conversion of the X-ray photons to OUV photons~\citep{Dai2018,Thomsen2022,Guolo2024}. Interestingly, the predicted SEDs always peak at a wavelength shorter than 1000~\AA\, although the optical-NUV segment can be naively fitted by a blackbody function with a temperature of $10^4$~K, aligning with observations~\citep{Gezari2021,vV2021,Hammerstein2023}. This approach could offer a promising solution to the "missing energy" puzzle, wherein the observed electromagnetic output of normal optical TDEs is at least one to two orders of magnitude lower than the anticipated energy released during the accretion of a star~\citep{Piran2015,Lu2018}. In addition, it is also possible that the SEDs of TDEs in AGNs differ greatly from those in normal galaxies due to the pre-existing AGN disk, as suggested by the simulations in \citet{Chan2021}. 

The integrated energy of PS16dtm derived from the observed OUV luminosity up to now (see bottom panel of Figure~\ref{dust}) is $1.86\times10^{52}$~erg. As introduced in the model fitting (\S\ref{modelfit}), the real \lbol\ has been underestimated by a factor of $\lambda$, and the same applies to the true energy. Assuming a typical 0.1~$\mu$m-sized silicate grain, $\lambda$ is 87.7 and the energy should then be as high as $\sim1.6\times10^{54}$~erg. It indicates that either the disrupted star is very massive, i.e. $>10$~\msun, or the energy output has been greatly enhanced by AGN (see discussion in \S\ref{AGNenhance}).  
Our results can also explain the X-ray dimming after the outburst of PS16dtm, with the low X-ray state persisting until the latest observations~\citep{Petrushevska2023}.  The current \lbol\ of $\sim10^{42.5}$~\lum, when multiplied by a factor of $\lambda=87.7$, yields a luminosity of $\sim2.8\times10^{44}$~\lum, that is still close to Eddington accretion. The X-ray emitting region is likely still obscured by the accretion-driven outflow in the reprocessing scenario.
However, we caution that our derivation of \lbol\ and total energy from the dust echo is subject to significant uncertainties due to insufficient information about the grain properties. Strictly, we cannot completely dismiss the possibility that the dust composition or size in PS16dtm is distinct. Future MIR spectroscopic observations with the James Webb Space Telescope (JWST) may refine our understanding of grain properties, leading to a more accurate estimate of \lbol.

\subsection{TDE energy output enhanced by AGN?}
\label{AGNenhance}

The energy budget for TDEs in AGNs is likely to be more complex than that of regular TDEs. To first order, AGNs offer a dense gas and dust environment at the subparsec scale around SMBHs, including the standard components of the accretion disk, broad-line region (BLR) gas, and dusty torus. In contrast, TDEs occurring in dormant SMBHs have a much more sparse environment, as evidenced by a notably weaker IR echo~\citep{Jiang2021b,vV2021b}. Is the energetic radiation of PS16dtm mainly driven by the environmental difference?

The luminous and long-lived dust echo of PS16dtm suggests an exceptionally high peak \lbol, corresponding to a highly super-Eddington accreting system. This implies that the circularization in this TDE is likely to be very rapid and efficient, with most of the stellar material being promptly accreted. Recent simulations have shown that stream-disk shocks can indeed effectively circularize returning debris~\citep{Steinberg2024}, which may be the case of PS16dtm, given the presence of a prior disk. Additionally, the high-speed outflow induced by super-Eddington accretion will inevitably collide with the BLR gas, including that which has been evaporated from the original inner torus. A broad blueshifted \mgii\ absorption feature, indicative of an outflow with a velocity of $>10^4$~\kms, has been detected in the UV spectrum of PS16dtm~\citep{Blanchard2017}. The outflow-BLR interaction creates a physical process very similar to that observed in Type IIn supernovae, where the interaction between ejecta and circumstellar material mainly powers its radiation. Assuming a typical outflow velocity of $\sim0.1c$, the ejecta kinetic energy is expected to be $10^{52}$~erg when 1\msun\ is ejected~\citep{Moriya2017}, indicating potentially a significant contribution to the total observed energy.

Therefore, the potential physical relationship between the energy output of a TDE and its environment may pose a limitation of the IR echo as a bolometer for characterizing the bulk properties of TDE energy, as it is biased towards more energetic events in gas/dust-rich environments that are conducive to producing detectable IR echoes. It is worth noting that some normal optical TDEs with clear IR echo detection, such as ASASSN-14li~\citep{Jiang2016,Prieto2016}, also exhibit tentative, albeit weak, AGN activity. Future missions with high-precision IR light curves could provide an opportunity to measure the energy of such targets more accurately and to test the environmental dependence of TDE energy output in a more quantitative manner.

\subsection{Prospect of Resolving the Giant Dusty torus}

It is interesting to explore the possibility of directly mapping such a giant dusty torus revealed by the dust echo of PS16dtm. At an angular distance of 313~Mpc ($z=0.08044$), the estimated inner torus radius of 1.59~pc subtends an angular size of 1.05 milliarcseconds, which is close to the resolution limit of the evolutionary NIR interferometry GRAVITY mounted on the Very Large Telescope Interferometer (VLTI)~\citep{GRAVITY2020,GRAVITY2024} and other conceptual facilities, i.e., MIR up-conversion stellar interferometer based on a synthetic long baseline~\citep{Han2024}. We have estimated the expected K-band magnitude to be 15, assuming blackbody dust emission. This magnitude is fainter than the current threshold brightness of GRAVITY ($K\approx10$). However, with the upgrade of GRAVITY to GRAVITY+, these limits will be significantly pushed further~\citep{Gravity+2022}.

In the future, it would be encouraging to compare the measurement from the dust echo model with that from direct imaging of the ring illuminated by a TDE such as PS16dtm. The prospects are quite promising in the GRAVITY+ era, which offers greatly improved sky coverage and enhanced capabilities for faint-science and high-contrast observations~\citep{Gravity+2022}. Meanwhile, hundreds to thousands of TDEs will be discovered by advanced surveys such as the Rubin Observatory Legacy Survey of Space and Time (LSST, \citealt{LSST2019,LSST2020}) and the Wide Field Survey Telescope (WFST, \citealt{Lin2022,Wang2023}). Their IR echoes can also be explored through IR time-domain surveys such as the NEO Surveyor~\citep{Mainzer2023}. A joint analysis of these observations could even aid in probing the Hubble constant, particularly through the measurement of the parallax distance to 3C 273 via spectroastrometry and reverberation mapping of its broad-line region~\citep{Wang2020}. However, we must acknowledge that the systematic uncertainties involved in the dust echo modeling are substantial (as illustrated in \S\ref{modelfit}), particularly when the type and size of the dust grains cannot be accurately determined. They may limit the applicability and reliability of the methodology and should be seriously considered in the future.

\section{Conclusion}

A luminous IR echo has been identified as a universal characteristic for TDEs in AGNs due to the presence of a dusty torus, but the long-lived echo in PS16dtm stands out as a unique case. 
After the initial discovery of a high temperature indicative of dust sublimation~\citep{Jiang2017}, its IR emission shows a high ($\sim10^{43.5}$~\lum) but slowly rising phase over 7 years, while the dust temperature remains relatively stable. A prolonged IR emission typically indicates a large dust structure that responds to transient TDE emission within the framework of the simplified dust echo model. Moreover, it suggests that the peak bolometric luminosity of the central outburst is exceptionally high, making the torus inner radius pushed outward. This motivates us to investigate this particular nuclear transient from the perspective of IR echo. However, the WISE/NEOWISE mission has recently ended, and regular MIR photometry will not be available for at least the next few years. Therefore, it is an opportune time to carry out a comprehensive study of the echo using its full MIR light curves.

In an effort to elucidate this remarkable and prolonged IR echo, we have devised a refined dust echo model that incorporates an expanding radius of the dust shell and considers the correlation between peak luminosity and radius. The model fit has yielded specific parameter values, including a dusty torus receding from 1.10 to 1.59 pc by evaporating the dust close to the BH. It is crucial to acknowledge the large inherent uncertainty in our simulations, highlighting the need for further observational tests by closely monitoring the evolution of the IR light curve in the upcoming years. The planned NEO Surveyor~\citep{Mainzer2023},  the successor to the NEOWISE mission, is expected to launch in 2027 and will allow us to test our model prediction. Alternatively, near-infrared (NIR) facilities can monitor the evolution in the coming years to bridge the time gap, albeit with less efficiency. However, some ground-based MIR facilities, such as MIRAC-5 on the Multiple Mirror Telescope~\citep{Bowens2022} and MIMIZUKU at the University of Tokyo Atacama Observatory~\citep{Kamizuka2020}, may also attempt to participate in long-term monitoring. With a continuous light curve extending into the decay phase, the torus structure can be much better constrained with an updated echo model.

Our echo model also unveils a remarkably high peak \lbol\ albeit dependent on the dust properties. Assuming a typical silicate grain with a size of $0.1\mu$m, the required \lbol\ is about $6\times10^{46}$~\lum, that is $>100$ times the Eddington luminosity for a $\sim10^6$~\msun\ in PS16dtm. Such a high \lbol\ could also explain the strange UV-deficient SED in an intrinsic power law form, as predicted by the reprocessing model~\citep{Dai2018}, but with a high dust extinction. Normal TDEs are known to have a so-called missing-energy puzzle, and the unobserved extreme-UV photons might account for the missing fraction. An alternative way to measure them is through their IR echoes, which has also been tentatively demonstrated by some previous work (e.g., \citealt{vV2016,Dou2017,Wang2022}). In PS16dtm, there is no missing energy at all, while the excess energy output may be enhanced by the pre-existing AGN. Thus, it needs to be emphasized that the IR echo method itself may introduce a selection effect, favoring TDEs in gas- or dust-rich environments, which could systematically produce more energetic events.
Detailed case studies of IR echoes for more TDEs in AGNs and a comprehensive comparison with normal TDEs are needed in the future.

Another interesting implication of our model is that the current giant inner dust radius ($\sim1.6$~pc) is close to the resolution limit of NIR interferometry instruments such as VLTI/GRAVITY. Although its $K$-band magnitude is much lower than the threshold brightness, it is still a promising topic to explore for future brighter or closer similar events, particularly in the era of the upgraded GRAVITY+. Therefore, we believe that the pc-scale dust environments of SMBHs, both active and quiescent, can be much better characterized using techniques of interferometry and TDE dust echo measurements.  \\

We sincerely thank the referee for very positive and constructive comments, which helped improve our manuscript significantly. N.J. dedicates this paper to the recently retired WISE/NEOWISE satellite, which has been with him throughout his research career over the past decade. This work is supported by the National SKA
Program of China (2022SKA0130102), the National Key Research and Development Program of China (2023YFA1608100), the National Natural Science Foundation of China (grants 12192221,12393814,12073025), the Strategic Priority Research Program of the Chinese Academy of Sciences (XDB0550200) and the China Manned Space Project. The authors gratefully acknowledge the support of the Cyrus Chun Ying Tang Foundations. 
This research makes use of data products from the Wide-field Infrared Survey Explorer, which is a joint project of the University of California, Los Angeles, and the Jet Propulsion Laboratory/California Institute of Technology, funded by the National Aeronautics and Space Administration. This research also makes use of data products from NEOWISE-R, which is a project of the Jet Propulsion Laboratory/California Institute of Technology, funded by the Planetary Science Division of the National Aeronautics and Space Administration. This research has made use of the NASA/IPAC Infrared Science Archive, which is operated by the California Institute of Technology, under contract with the National Aeronautics and Space Administration.

\end{document}